# Inter-valence charge transfer and charge transport in the spinel ferrite ferromagnetic semiconductor Ru-doped CoFe$_2$O$_4$

Masaki Kobayashi[1,2,*], Munetoshi Seki[1,2], Masahiro Suzuki[3], Miho Kitamura[5], Koji Horiba[5], Hiroshi Kumigashira[5,6], Atsushi Fujimori[3,4], Masaaki Tanaka[1,2], and Hitoshi Tabata[1,2]

[1]*Department of Electrical Engineering and Information Systems, The University of Tokyo, 7-3-1 Hongo, Bunkyo-ku, Tokyo 113-8656, Japan*

[2]*Center for Spintronics Research Network, The University of Tokyo, 7-3-1 Hongo, Bunkyo-ku, Tokyo 113-8656, Japan*

[3]*Department of Physics, The University of Tokyo, 7-3-1 Hongo, Bunkyo-ku, Tokyo 113-0033, Japan*

[4]*Department of Applied Physics, Waseda University, Okubo, Shinjuku, Tokyo 169-8555, Japan*

[5]*Photon Factory, Institute of Materials Structure Science, High Energy Accelerator Research Organization (KEK), 1-1 Oho, Tsukuba 305-0801, Japan*

[6]*Insitute of Multidisciplinary Research for Advanced Materials (IMRAM), Tohoku University, Sendai 980–8577, Japan*

*Email: masaki.kobayashi@ee.t.u-tokyo.ac.jp

**Abstract**

Inter-valence charge transfer (IVCT) is electron transfer between two metal $M$ sites differing in oxidation states through a bridging ligand: $M^{n+1} + M'^{m} \rightarrow M^{n} + M'^{m+1}$. It is considered that IVCT is related to the hopping probability of electron (or the electron mobility) in solids. Since controlling the conductivity of ferromagnetic semiconductors (FMSs) is a key subject for the development of spintronic device applications, the manipulation of the conductivity through IVCT may become a new approach of band engineering in FMSs. In Ru-doped cobalt ferrite CoFe$_2$O$_4$ (CFO) that shows ferrimagnetism and semiconducting transport properties, the reduction of the electric resistivity is attributed to both the carrier doping caused by the Ru substitution for Co and the increase of the carrier mobility due to hybridization between the wide Ru 4$d$ and the Fe 3$d$ orbitals. The latter is the so-called IVCT mechanism that is charge transfer between the mixed valence Fe$^{2+}$/Fe$^{3+}$ states facilitated by bridging Ru 4$d$ orbital: Fe$^{2+}$ + Ru$^{4+}$ $\leftrightarrow$ Fe$^{3+}$ + Ru$^{3+}$. To elucidate the emergence of the IVCT state, we have conducted x-ray absorption spectroscopy (XAS) and resonant photoemission spectroscopy (RPES) measurements on non-doped CFO and Co$_{0.5}$Ru$_{0.5}$Fe$_2$O$_4$ (CRFO) thin films. The observations of the XAS and RPES spectra indicate that the presence of the mixed valence $Fe^{2+}/Fe^{3+}$ state and the hybridization between the Fe $3d$ and Ru $4d$ states in the valence band. These results provide experimental evidence for the IVCT state in CRFO, demonstrating a novel mechanism that controls the electron mobility through hybridization between the $3d$ transition-metal cations with intervening $4d$ states.

**I. Introduction**

Inter-valence charge transfer (IVCT) occurs in mixed-valence coordination complexes, and is an electron transfer between two metal $M$ sites differing in oxidation states through a bridging ligand: $M^{n+1} + M'^{m} \rightarrow M^{n} + M'^{m+1}$, and is usually related to photoexcitation phenomena in these compounds [1, 2]. Additionally, magneto-optical properties of ferrites or iron oxides that seem to originate from IVCT such as magneto-optical Kerr rotation and photo-induced magnetization have been reported so far [3, 4, 5, 6, 7, 8]. In semiconducting or insulating materials with mixed-valence states, especially for metal-organic frameworks, IVCT is considered to be related to the transport properties with hopping conduction [6, 9, 10, 11, 12, 13]. Then, it is expected that one can control magneto-optical and transport properties of mixed-valence semiconductors or insulators through IVCT.

Ferromagnetic semiconductors (FMSs) having both semiconducting and ferromagnetic properties are key materials for spintronics, which is a research field to use both the charge and spin degrees of freedom for functional electronic devices [14, 15, 16]. Controlling the conductivity of FMSs with keeping their ferromagnetic properties is useful for utilizing FMSs to spintronics devices such as magnetic random-access memories (MRAMs) and spin transistors. Spinel ferrites $M$Fe$_2$O$_4$ [$M$ = 3$d$ transition metals (TMs)] are promising magnetic materials for spintronics because of their chemical stability and high Curie temperatures [17]. Indeed, spinel ferrite layers in magnetic tunnel junctions have been demonstrated to act as spin filters [18, 19]. CoFe$_2$O$_4$ (CFO) with the inverse spinel structure shows ferrimagnetism with the Curie temperature ($T_C$) of 793 K and CFO epitaxial thin films have been studied for the application to spintronics devices [10, 20, 21, 22, 23]. In CFO, the stoichiometric valence states of the constituent Co and Fe ions are $Co^{2+}$ at the octahedral crystal-field ($O_h$) site and $Fe^{3+}$ at both the $O_h$ and the tetrahedral crystal-field ($T_d$) sites, as shown in Fig. 1. In contrast to the semiconducting CFO, magnetite $Fe_3O_4$ shows metallic transport property because of double-exchange interaction between the $Fe^{2+}$ and $Fe^{3+}$ sites. Recently, Iwamoto *et al.* [10] have succeeded in increasing the conductivity of $CoFe_2O_4$ by Ru doping, where the $Ru^{4+}$ ions are considered to preferentially occupy the $Co^{2+}$ sites in CFO [7, 10]. Figure 2 shows the transport properties of Ru-doped $CoFe_2O_4$ thin films [10]. The results indicate that the Ru doping increases the carrier density and the electron mobility. The doped $Ru^{4+}$ ions in CFO are expected to act as double donors and that the Ru doping increases not only the carrier concentration but also the hopping probability though hybridization between the Fe $3d$ and Ru $4d$ orbitals. It is considered that the carriers supplied by Ru doping generate the mixed valence state of $Fe^{2+}/Fe^{3+}$ in CFO, leading to the improvement of the conductivity through double-exchange interaction like $Fe_3O_4$. First-principles band calculation suggests that the increase of the hopping conduction occurs through hybridization between the Fe $3d$ and Ru $4d$ orbitals following the IVCT mechanism $Fe^{2+} + Ru^{4+} \leftrightarrow Fe^{3+} + Ru^{3+}$.

To prove the existence of such an IVCT state in CRFO, two key factors should be confirmed experimentally: (1) the $Fe^{2+}/Fe^{3+}$ mixed-valence state and (2) the hybridization between the Fe $3d$ and Ru $4d$ orbitals. In this letter, we have conducted x-ray absorption spectroscopy (XAS) and resonance photoemission spectroscopy (RPES) studies of non-doped CFO and Ru-doped $CoFe_2O_4$ thin films in order to obtain experimental evidence for the IVCT state in CRFO. XAS and RPES enable us to probe element- and orbital-specific electronic structures. The XAS spectra reveal the valence states of Fe and Co in CRFO. The RPES spectra indicate that the partial density of states (PDOS) of Fe $3d$ change with Ru doping. These experimental findings suggest that the IVCT state is realized in Ru-doped CFO. The manipulation of the conductivity through IVCT may become a new approach of band engineering in FMS materials.

**II. Experimental**

$CoFe_2O_4$ (CFO) and $Co_{0.5}Ru_{0.5}Fe_2O_4$ (CRFO) epitaxial thin films were grown on single crystal $\alpha$-$Al_2O_3$(0001) substrates using the pulsed laser deposition (PLD) technique with an ArF-excimer laser with the wavelength was 193 nm, the frequency was 5 Hz, and the fluence was $E = 60$ mJ. The details of the thin-film growth are described elsewhere [10]. The PES and XAS measurements were performed at BL-2A of Photon Factory (PF), High Energy Acceleration Research Organization (KEK) [24]. The total energy resolution was set to be 100-250 meV for PES measurements using photon energy of 400-1200 eV. The binding energies were calibrated by measuring the $E_F$ of a gold foil which was electrically contacted to the samples. The PES measurements were conducted with an SES2002 electron analyzer at room temperature under the base pressure below $2.0 \times 10^{-8}$ Pa. The XAS spectra were measured in the total electron yield (TEY) mode.

**III. Results and discussion**

Basically, XAS spectra reflect the local electronic structures of specific elements. We have performed XAS measurements on CRFO thin films to elucidate the valence state of each element. Figure 3 shows the Co $L_{2,3}$ XAS spectra of parent CFO and Ru-doped one. The spectra of CFO and CRFO show multiplet structures that are different between them. Comparing the observed Co $L_{2,3}$ spectra with those of reference compounds CoO ($Co^{2+}$ $O_h$) and LaCoO$_3$ ($Co^{3+}$ $O_h$) [25], the spectrum of CRFO is similar to that of CoO and the spectrum of CFO seems to be a superposition of them. Here, the $Co^{3+}$ $O_h$ spectrum represents cation inversion or Co-antisite defects (the inter-site cation exchange between the A $T_d$ and B $O_h$ sites, and the on-site cation replacement at the A sites) because the XAS spectra of the $Co^{3+}$ $O_h$ is similar to that of the $Co^{3+}$ $T_d$ [26]. Here, the cation inversion defects is represented by the inversion parameter $y$ defined by the chemical formula [Fe$_{1-y}$Co$_y$]$_{Td}$[Fe$_{1+y}$Co$_{1-y}$]$_{Oh}$O$_4$. To estimate quantitatively the ratio of these oxidation (valence) states, the Co $L_{2,3}$ XAS spectra of CFO and CRFO are fitted by a linear combination of the reference spectra of CoO and LaCoO$_3$. Figures 3(b) and 3(c) show such decomposition analyses for the Co $L_{2,3}$ XAS spectra. Figure 3(b) shows that the Co $L_{2,3}$ XAS spectrum of CRFO is fitted by the spectrum of CoO and that there is nearly no contribution of the $Co^{3+}$ state to the experimental spectrum within the accuracy of the analysis. In contrast, the Co $L_{2,3}$ XAS spectrum of CFO is decomposed into the ~75% $Co^{2+}$ and the ~25% $Co^{3+}$ components, as shown in Fig. 3(c), indicating that part of the $Co^{2+}$ ions at the B site becomes $Co^{3+}$ at the B site or move to the A site. The latter case means the existence of antisite defects, namely, Co ions at the A site and excess Fe ions at the B site. This is consistent with that non-negligible cation inversion defects usually exist in spinel ferrite thin films [22, 23]. In CRFO, on the other hand, electron-rich environment induced by Ru doping suppresses the occurrence of $Co^{3+}$ ions and possibly the cation inversion. It should note here that the $Co^{2+}$ components possibly include both the $Co^{2+}$ at the B site and the antisite defect of the $Co^{2+}$ at the A site. The chemical compositions are discussed below considering the decomposition analysis of the Fe $L_{2,3}$ XAS spectra.

Figure 4 shows the Fe $L_{2,3}$ XAS spectra of the parent CFO and CRFO thin films. The Fe $L_{2,3}$ XAS spectra show multiplet structures and the spectral line shapes are similar to that of $Fe_2O_3$, as shown in Fig. 4(a). In contrast to the Co $L_{2,3}$ XAS spectra, the Fe XAS spectrum of CRFO film looks nearly identical to that of CFO one. To elucidate the Fe components precisely, the Fe $L_{2,3}$ XAS spectra are decomposed into various valence and crystal-field states using linear combinations of reference spectra of $Fe_2O_3$ ($Fe^{3+}$ $O_h$) [27], $YBaCo_3FeO_7$ ($Fe^{3+}$ $T_d$) [28], and FeO ($Fe^{2+}$ $O_h$) [29]. As show in Figs. 4(b) and 4(c), the linear combinations of the reference spectra well reproduce the experimental spectra of CRFO and CFO samples. The ratios of the components are listed in Table I. As expected from the line shapes, the ratio among the Fe components of CRFO is similar to that of CFO. Considering the possible existence of the cation inversion defects in the CFO thin film observed in the Co $L_{2,3}$ XAS spectrum, the $Fe^{2+}$ component in the CFO film likely originates from the Fe-antisite defect, where the excess Fe ions substitute for $Co^{2+}$ ions at the $O_h$ sites, as in the case for the $Co^{3+}$ component tentatively assigned to cation inversion defects [Fig. 3(c)]. Based on the analysis, the chemical composition of CFO is estimated as $[Fe_{0.67}Co_{0.33}]_{Td}[Fe_{1.33}Co_{0.67}]_{Oh}O_4$ ($y = 0.33$). On the other hand, the presence of the $Fe^{2+}$ component in the CRFO film would be induced by the electron doping caused by the Ru substitution because there will be few cation inversion defects in the CRFO film. Although the ratio of $Fe^{2+}$ is expected to be 25% in CRFO from the nominal stoichiometry assuming $Ru^{4+}$, the deduced value is ~10.5%. Since there is nearly no $Co^{3+}$ component in CRFO, this quantitative difference may come from Ru-antisite defects substituting for the Fe sites. These results provide spectroscopic evidence for the $Fe^{2+}/Fe^{3+}$ mixed-valence state in the CRFO thin film caused by the Ru doping, even though there is a quantitative difference.

To elucidate the possible hybridization between the Fe $3d$ and Ru $4d$ orbitals, which is another key factor for the IVCT, the Fe and Co $3d$ PDOS in the valence band (VB) that arise from the different valence states have been obtained using RPES. Figure 5 shows the RPES spectra of the CFO and CRFO thin films. When incident photon energy $h\nu$ corresponds to that for the core-hole excitation, the photoemission intensity of the orbitals into which the excited electron orbital is resonantly enhanced. Difference between the on-resonance spectrum and off-resonant spectrum reflects the 3$d$ PDOS corresponding to those orbitals. Since the 2$p$-3$d$ excitation energy for each component is different, one can obtain the PDOS dominated by each component by tuning $h\nu$ (it is difficult to separate these PDOS perfectly because of the overlap of the XAS spectra between these components). For instance, as shown in Fig. 5(a), the $Fe^{2+}$ and $Fe^{3+}$ PDOS of the thin films are obtained from the on-resonance spectra measured at $h\nu$ of 708.5 eV and 710 eV, respectively. Figure 5(b) shows the Fe and Co 3$d$ PDOS and off-resonance spectra of CFO and CRFO films. Here, the on-resonance spectra are taken at the core-hole excitation energies for the $Co^{2+}$, $Co^{3+}$, $Fe^{2+}$, and $Fe^{3+}$ states, and the spectra are normalized to the spectral area. For the Co-3$d$ PDOS, the differences of the spectra between the samples reflect the effect of Ru substitution for the Co sites. Since the Co 3$d$ spectral intensity near the valence-band maximum (VBM) decreases with Ru doping, we conclude that the Co 3$d$ PDOS hardly contributes to the increase of the conductivity. Similarly, the $Fe^{3+}$ PDOS shows nearly the same spectral changes as Co 3$d$ with Ru doping, suggesting few direct contributions of the $Fe^{3+}$ component to the improvement of the conductivity.

It should note here that the differences of the Co off-resonance spectra between the CFO and CRFO films reflect the appearance of the Ru 4$d$ PDOS and the decrease of the Co-3$d$ PDOS with Ru doping. Considering the cross section of the atomic orbitals around this incident-photon energy region, the VB spectra without resonance mainly reflect the Ru 4$d$ PDOS [30]. Since the spectral intensity near the VBM increases with Ru doping, the Ru 4$d$ PDOS predominantly contributes to the DOS near the VBM, qualitatively consistent with the increase of the conductivity with Ru doping. To reveal the 3$d$ component hybridized with the Ru 4$d$ orbitals, the additional 3$d$ PDOS near VBM induced by Ru doing are examined. Figure 5(c) shows comparison of the various PDOS in the vicinity of VBM that have been normalized to the intensity at $E_B$~1.5 eV. The difference in the off-resonance spectra indicates that the Ru 4$d$ PDOS is located near the VBM. Note also that additional $Fe^{2+}$ PDOS appears with Ru doping and the position of the $Fe^{2+}$ PDOS is nearly identical to that of the Ru 4$d$ PDOS [see arrows in Fig. 5(c)]. In contrast, the $Co^{3+}$, $Co^{2+}$, and $Fe^{3+}$ PDOS have the same slopes irrespective to the Ru doping. These observations are consistent with the first-principle calculation that the B-site Fe 3$d$ orbitals hybridizes with the Ru 4$d$ orbital except for the opening of the conductivity gap in experiment [10]. The band gap may come from Coulomb interaction, i.e., a Mott gap. Thus, the present result is consistent with hybridization between the Fe 3$d$ and Ru 4$d$ orbitals in CRFO.

The present findings, that is, the mixed-valence state of $Fe^{2+}/Fe^{3+}$ and the hybridization between the Fe 3$d$ and Ru 4$d$ orbitals, provide spectroscopic evidence for the IVCT state in Ru-doped CFO. In CRFO, the O 2$p$ orbitals that are located at the nearest neighbor atoms of the B sites make bonding states directly with the Fe 3$d$ $e_g$ and Ru 4$d$ $e_g$ orbitals. It is likely that the ligand O 2$p$ orbitals bridge electron hopping between the Fe and Ru sites such as $Fe^{2+} + Ru^{4+} \leftrightarrow Fe^{3+} + Ru^{3+}$. Here, while IVCT for photoexcitation is usually expressed with a single-headed arrow between $M$ atoms, we use a double-headed arrow to IVCT for transport. In contrast to the $M$-$M'$ distances for IVCT in coordination complexes reportedly in the range of 10~25 Å [2], the effective distance for the IVCT possibly becomes longer in single crystalline TM compounds like spinel ferrite oxides because the ligand bands bridging the IVCT are expanded the entire crystal. Compared with the $M$-$M'$ distance in coordination complexes, in which the photo-excitation induces charge transfer in the environment of highly insulating nature, the longer effective distances for the IVCT in the TM compounds lead to the increase of the hopping probability (or electron mobility) through IVCT as a consequence of the reduction of energy barrier for the charge transfer. This may explain the increase of the conductivity by Rh doping in $Fe_2O_3$ ($Fe_{1.8}Rh_{0.2}O_3$) without mixed-valence states [31], where the doped $Rh^{3+}$ ions isovalently substitute for the $Fe^{3+}$ sites. The same mechanism may also explain the increase of the hopping probability through IVCT of $FeTiO_3$ ilmenite [9], $Fe^{2+} + Ti^{4+}$ $\leftrightarrow Fe^{3+} + Ti^{3+}$, without double-exchange interaction [since the $Ti^{4+}$ ion ($d^0$ configuration) is non-magnetic]. As in the cases for $Fe_3O_4$ and $(La,Sr)MnO_3$, double-exchange interaction between magnetic ions with the mixed-valence states stabilizes the ferromagnetism and contributes to the increase the metallic conductivity [32]. In contrast, the conduction mechanism through IVCT in solids will be applicable for ferromagnetic semiconducting materials having low carrier concentrations with hopping conduction. Furthermore, as described in the introduction, the IVCT has the capability to modify the optical properties of semiconducting materials. It follows from the above arguments that control of the IVCT in FMSs with mixed-valence states will be a new approach manipulating the electron mobility and magneto-optical properties through hybridization between the TM $d$ orbitals with maintaining the ferromagnetism.

**IV. Summary**

In this study, we have investigated the electronic structure of Ru-doped $CoFe_2O_4$ using XAS and RPES to elucidate the emergence of IVCT in semiconducting crystal. The Co $L_{2,3}$ XAS spectra suggest that although the 25% of the total amount of Co atoms exist as $Co^{3+}$ in the CFO thin film, there are only $Co^{2+}$ ions in the Ru-doped CFO thin film. The result that there is almost no $Co^{3+}$ ions in the Ru-doped CFO thin film provides an important aspect to control defect level for materials growth of spinel ferrites. The Fe $L_{2,3}$ XAS spectra indicate that the $Fe^{2+}/Fe^{3+}$ mixed-valence state is realized in CRFO. The observation using RPES demonstrates that the Ru $4d$ PDOS appears near the VBM and hybridizes with the $Fe^{2+}$ $3d$ state, and not with the Co ones. These findings provide spectroscopic evidence for the IVCT state between the Ru $4d$ and Fe $3d$ orbitals. Controlling the IVCT through the hybridization will open a new way to manipulate both the magneto-optical properties and the carrier mobility of FMSs.

**Acknowledgment**

This work was supported by a Grant-in-Aid for Scientific Research (Nos. 15H02109,16H02115, 17H04922, 18H05345) and Core-to-Core Program from the Japan Society for the Promotion of Science (JSPS), CREST (JPMJCR1777), and the MEXT Elements Strategy Initiative to Form Core Research Center. This work was partially supported the Spintronics Research Network of Japan (Spin-RNJ) and Basic Research Grant (Hybrid AI) of Institute for AI and Beyond for the University of Tokyo. This work at KEK-PF was performed under the approval of the Program Advisory Committee (Proposals 2015S2-005 and 2018G114) at the Institute of Materials Structure Science at KEK. A.F. is an adjunct member of Center for Spintronics Research Network (CSRN), the University of Tokyo, under Spin-RNJ.

**Table**

Table I. Decomposition analysis for the Co $L_{2,3}$ and Fe $L_{2,3}$ XAS spectra. It should note here that the $Co^{2+}$ components include both the $Co^{2+}$ at the B site and the possible antisite defect of the $Co^{2+}$ at the A site.

|  | A($T_d$) site | | B($O_h$) site | | |
| --- | --- | --- | --- | --- | --- |
|  | $Co^{3+}$ | $Fe^{3+}$ | $Co^{2+}$ | $Fe^{3+}$ | $Fe^{2+}$ |
| $CoFe_2O_4$ | 25% | 33.6% | 75% | 57.9% | 8.5% |
| $Co_{0.5}Ru_{0.5}Fe_2O_4$ | 0% | 37.2% | 100% | 52.3% | 10.5% |

**Figures**

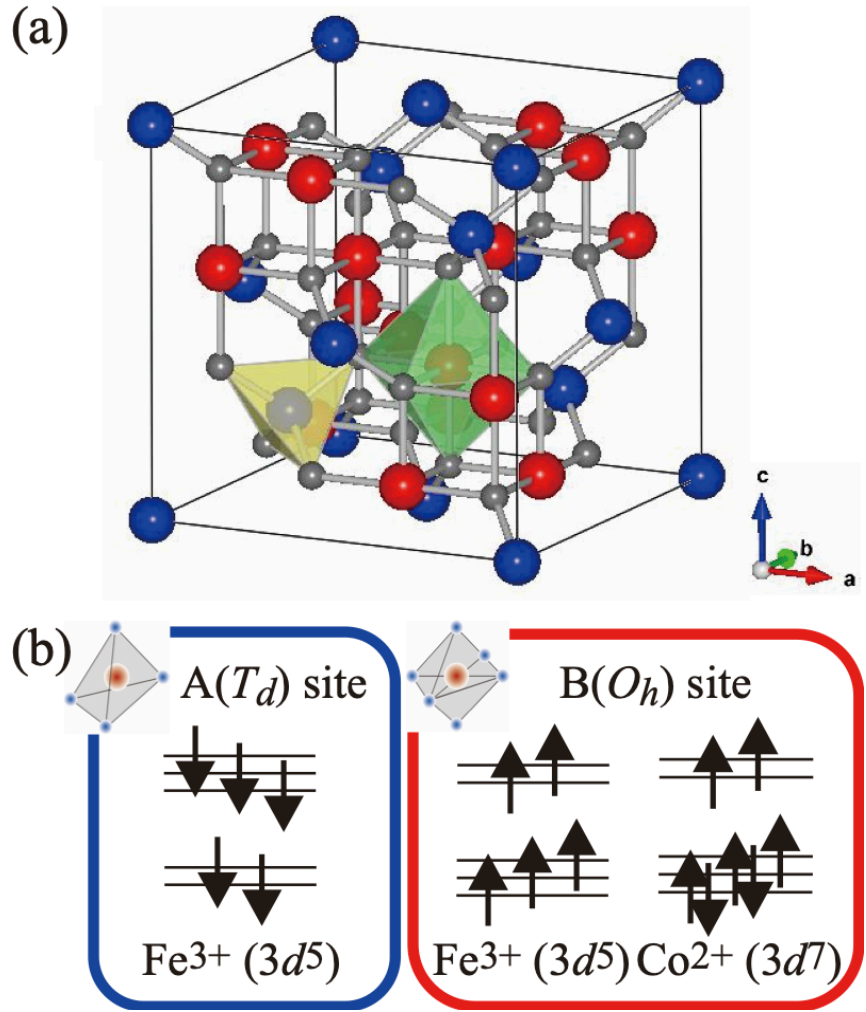

FIG 1: Crystal structure of inverse spinel $CoFe_2O_4$. (a) Unit cell of $CoFe_2O_4$. The magnetic sublattice of the A($T_d$) site antiferromagnetically couples with that of the B($O_h$) site. (b) Spin configurations of the A and B sites.

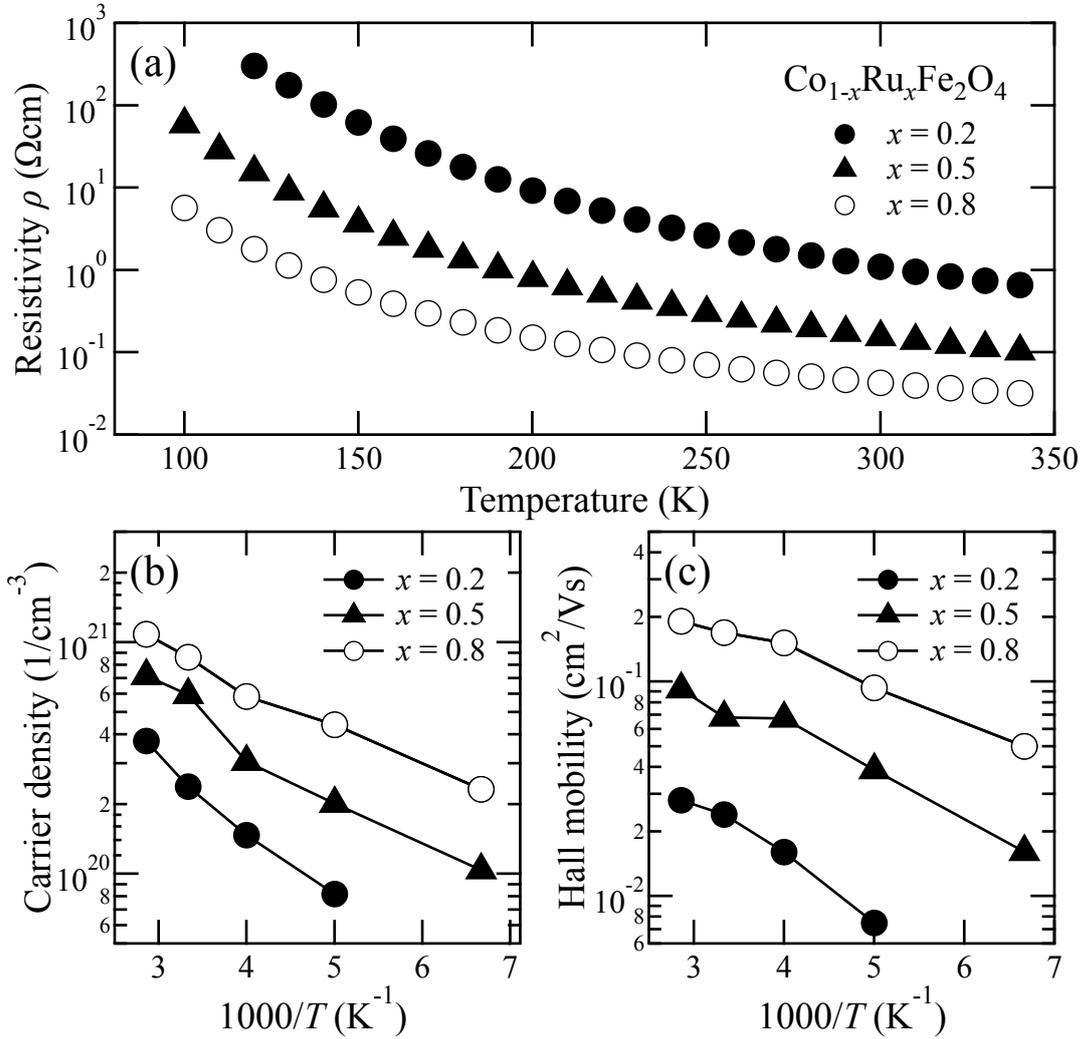

FIG. 2: Transport properties of $Co_{1-x}Ru_xFe_2O_4$ thin films [10]. (a) Temperature ($T$) dependence of resistivity $\rho$. (b), (c) Compositional dependence of the carrier density and the Hall mobility, respectively.

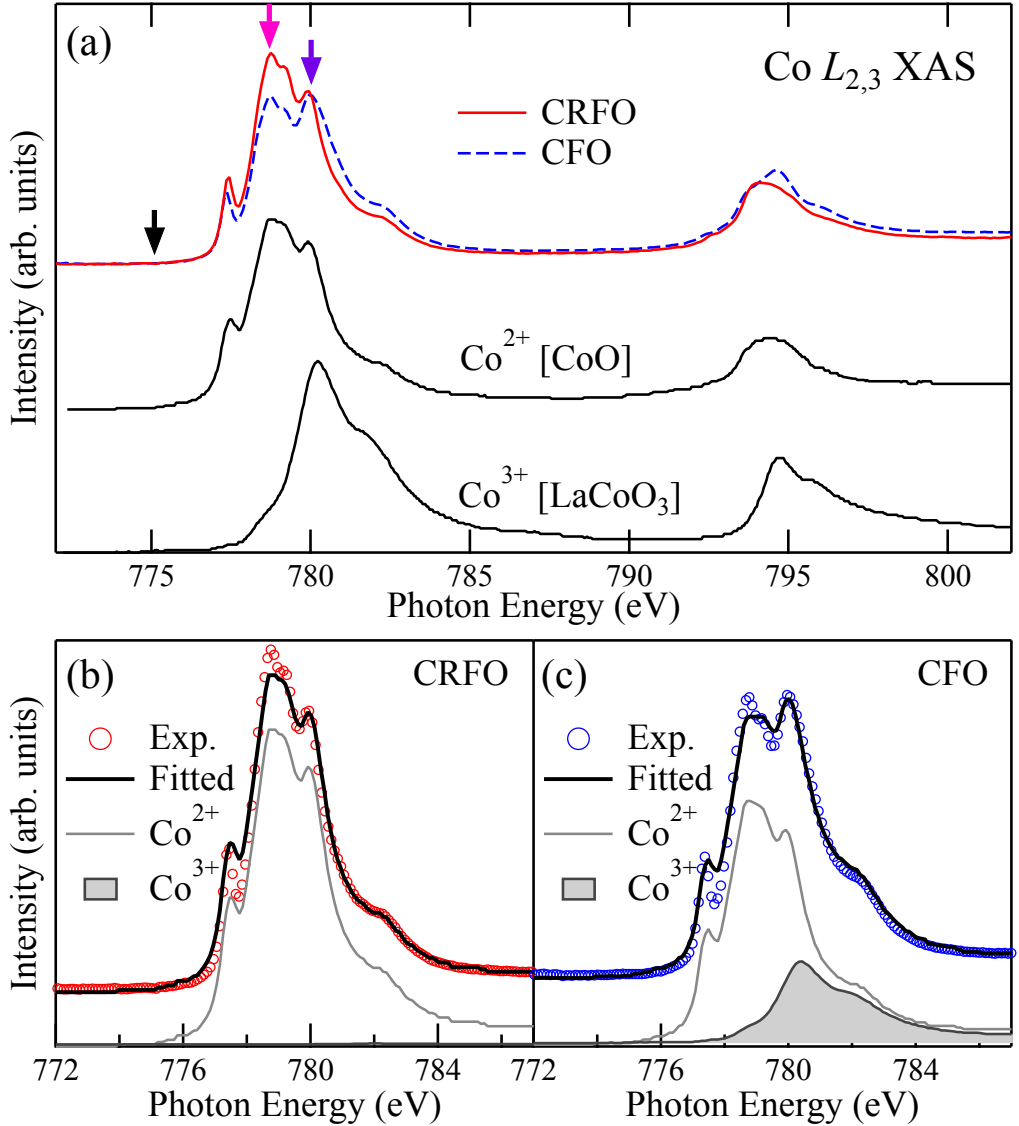

FIG. 3: Co $L_{2,3}$ XAS spectra of $Co_{1-x}Ru_xFe_2O_4$ thin films. (a) The XAS spectra of parent CFO and CRFO thin films. The spectra of CoO ($Co^{2+}$) and $LaCoO_3$ ($Co^{3+}$) are also shown as references [25]. The arrows denote excitation energies for RPES, i.e., 775 eV for the off-resonance, 778.7 eV for the $Co^{2+}$ on-resonance, and 780 eV for the $Co^{3+}$ on-resonance. (b), (c) Decomposition of the Co $L_{2,3}$ spectra for the CRFO and CFO thin films, respectively.

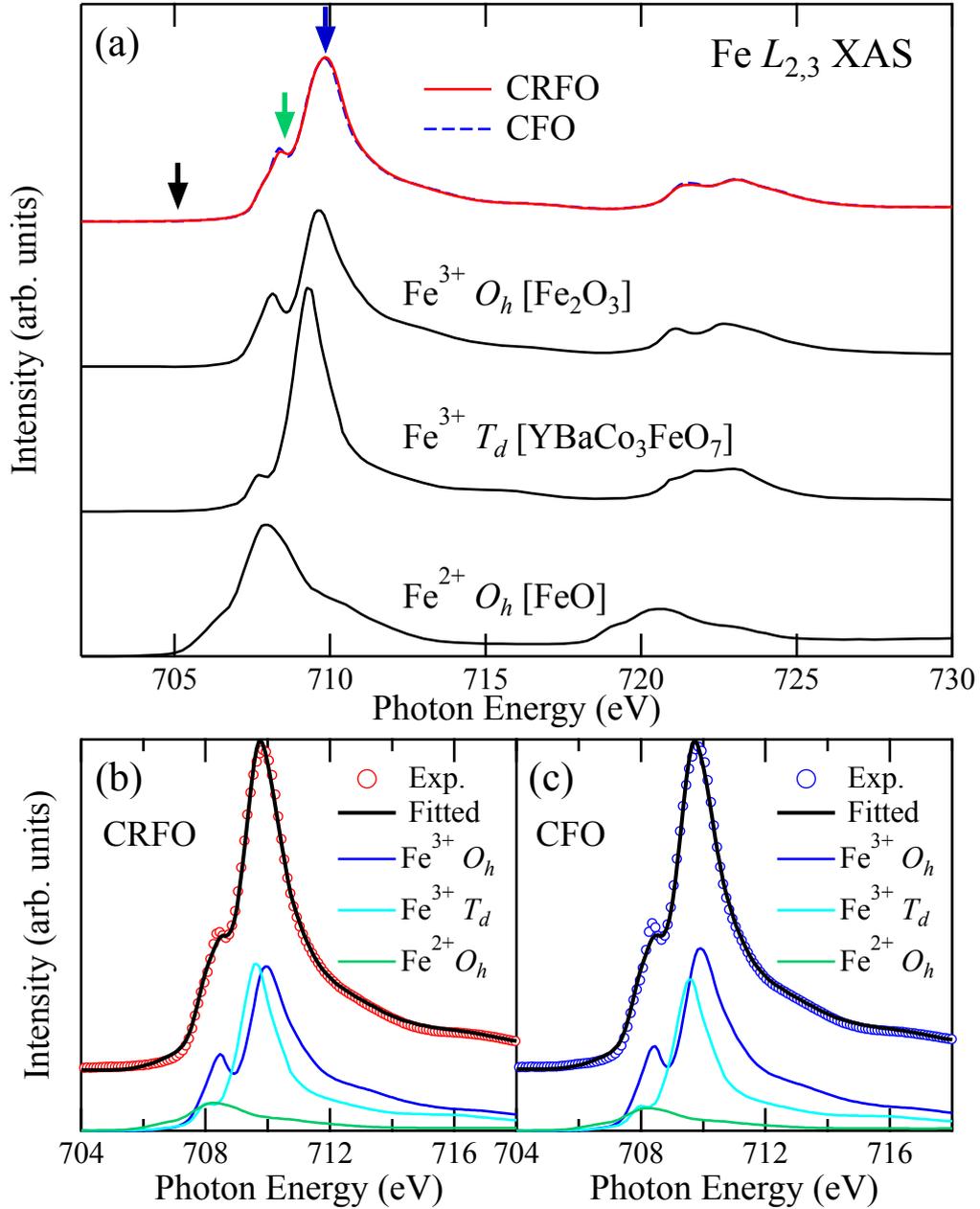

FIG. 4: Fe $L_{2,3}$ XAS spectra of $Co_{1-x}Ru_xFe_2O_4$ thin films. (a) The XAS spectra of parent CFO and CRFO thin films. The reference spectra of $Fe_2O_3$ ($Fe^{3+}$ $O_h$) [27], $YBaCo_3FeO_7$ ($Fe^{3+}$ $T_d$) [Hollmann_PRB_09], and FeO ($Fe^{2+}$ $O_h$) [29] are also shown. The arrows denote excitation energies for RPES, i.e., 705 eV for the off-resonance, 708.5 eV for the $Fe^{2+}$ on-resonance, and 710 eV for the $Fe^{3+}$ on-resonance. (b), (c) Decomposition of the Fe $L_{2,3}$ spectra for the CRFO and CFO thin films, respectively.

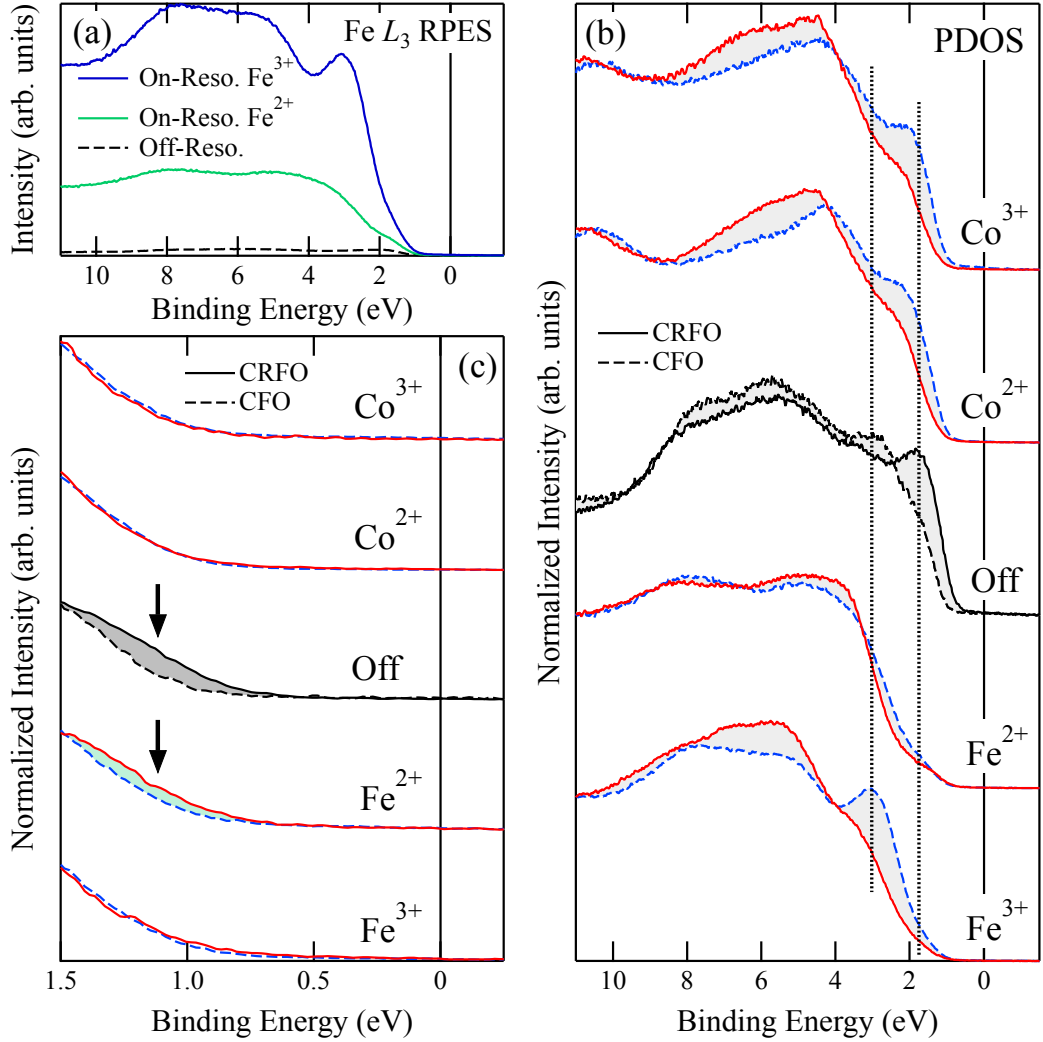

FIG. 5: Resonant photoemission spectra taken at Fe $L_3$ and Co $L_3$ edges of $Co_{1-x}Ru_xFe_2O_4$ thin films. (a) On- and off-resonance spectra at the Fe $L_3$ edge of CFO. (b) Partial density of states of the Fe $3d$ and Co $3d$ states. The shaded areas are difference of PDOS between CFO and CRFO films. The spectra have been normalized to the intensity integrated from $E_F$ to $E_B \sim 11$ eV. (c) Fe $3d$ and Co $3d$ PDOS near the valence band maximum. The spectra have been normalized to the intensity at $E_B = 1.5$ eV in order to emphasizing the spectral change near the VBM.